\documentstyle[aps,prl,epsfig,multicol]{revtex}
\begin{document}
\title{Three-Dimensional Superconductivity in the Infinite-Layer Compound 
       Sr$_{0.9}$La$_{0.1}$CuO$_2$ in Entire Region below $T_c$}
\author{Mun-Seog Kim, C. U. Jung, J. Y. Kim, Jae-Hyuk Choi, and Sung-Ik Lee}
\address{National Creative Research Initiative Center for Superconductivity
         and Department of Physics, Pohang University of Science and
         Technology, Pohang 790-784, Republic of Korea}
\date{\today}
\maketitle
\begin{abstract}
The infinite-layer compound ACuO$_{2}$ (A $=$ alkaline-earth ions) is regarded 
as the most suitable material for exploring the fundamental nature of the CuO$_2$ 
plane because it does not contain a charge-reservoir block, such as a rock-salt 
or a fluorite like block. We report that superconductivity in the infinite-layer
compound Sr$_{0.9}$La$_{0.1}$CuO$_2$ is of a three-dimensional nature, in contrast 
to the quasi two-dimensional superconducting behavior of all other cuprates. The key 
observation is that the $c$-axis coherence length is longer than the $c$-axis 
lattice constant even at zero temperature. This means that the superconducting 
order parameter of one CuO$_{2}$ plane overlaps with those of neighboring CuO$_{2}$ 
planes all the temperatures below the $T_c$. Among all cuprates, only the 
infinite-layer superconductor shows such a feature.
\end{abstract}

\begin{multicols}{2}
The key ingredient of high-temperature superconductors (HTSC) is the CuO$_2$ 
plane in which superconductivity occurs. Besides the CuO$_2$ plane, the unit 
cell of HTSC generally contains a charge-reservoir block (CRB) which supplies 
holes or electrons into the conducting layer. However, the function of the CRB, 
beyond supplying carriers in the materials, is not yet completely clear.
In one respect, the CRB might simply be a spacer between the CuO$_2$ planes. 
In this case, the block reduces the layer-by-layer coupling. The strong 
anisotropic nature of HTSC is believed to be a reflection of this weak
interlayer coupling. In this context, the infinite-layer compounds 
ACuO$_{2}$ (A $=$ alkaline-earth ions) are remarkable,
because they do not have a CRB, so the simplicity of their crystal structure
may allow in-depth insight into the basic mechanism of cuprate superconductivity. 
Due to the absence of the CRB, infinite-layer compounds have two notable features. 
First, the distance from one unit cell to the next is the {\em shortest} among all the
cuprates.\cite{siegrist1} Secondly, as a matter of course, charge carriers can 
be supplied only from the cations at the A sites. For example, the carrier density 
of the stoichometric infinite-layer SrCuO$_2$ is zero. The partial substitution of 
La$^{+3}$ (or Nd$^{+3}$) for Sr$^{+2}$ results in superconductivity in these
compounds. \cite{smith1} In this case, the carriers are not holes, 
but electrons.\cite{jorgensen1,er1,jones1,liu1}

Recently, we successfully synthesized 
pure-phase Sr$_{0.9}$La$_{0.1}$CuO$_2$ (Sr(La)-112) with $T_c\simeq43$ K by using 
a cubic multi-anvil press. The details of the sample preparation will be given 
elsewhere.\cite{cujung1} In this work,  we measured the reversible magnetization 
as a function of the temperature and the angle between the $c$ axis and
the applied magnetic field. From analysis of the data, we found that the usual HTSC
two-dimensional (2D) temperature region of $\xi_c(T)<c$ below $T_c$ did not exist 
in this compound. This peculiar feature has not been observed in any other high-$T_c$ 
material. Until now, YBa$_2$Cu$_3$O$_{7-\delta}$ (Y-123) has been known to have the most 
strong interlayer coupling, but the three-dimensional (3D) temperature region is 
limited to near $T_c$ only. Previously,\cite{kitazawa1} it was claimed that the 
high-$T_c$ superconductivity occurred only on the 2D network of CuO$_2$ planes, since a 
3D network did not permit spin or charge fluctuation while a 1D structure did 
not establish long range order of superconductivity. Hence, 
our observation of 3D superconductivity in entire region below $T_c$ in Sr(La)-112 
provides a new aspect of high-temperature superconductivity.

\begin{figure}
\center
\includegraphics[width=7cm,height=5.5cm]{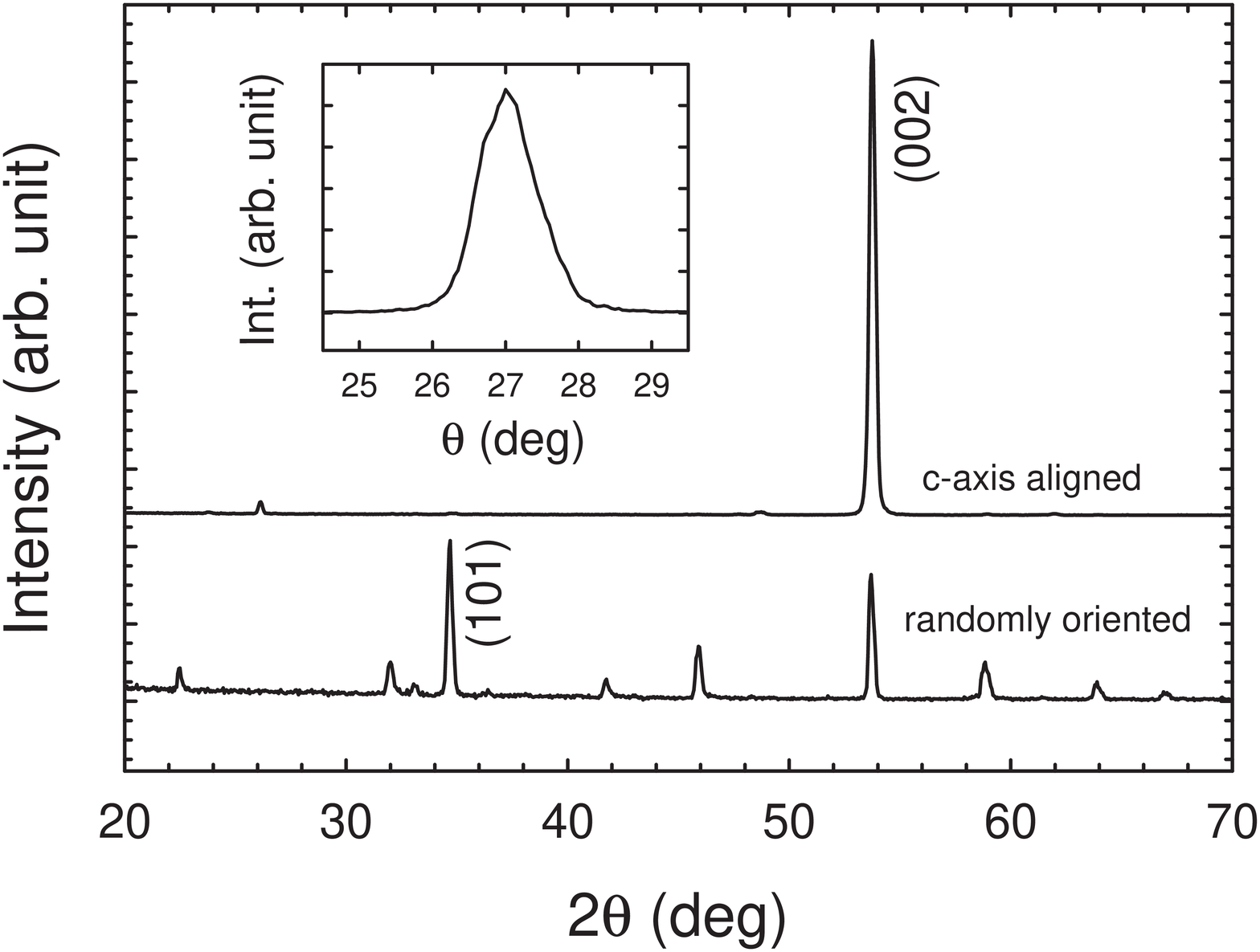}
\caption{XRD patterns of Sr$_{0.9}$La$_{0.1}$CuO$_2$ before and after the grain
         alignment. The inset shows x-ray rocking curve of the (002) reflection of
         aligned sample.}
\label{fig1}
\end{figure}

For the magnetization measurements, we aligned the grains of the sample 
in commercial epoxy under an external magnetic field 
of 11 T.\cite{Farrell1,mskim30,mskim21,mskim15,mskim14} 
Fig.\ \ref{fig1} displays the x-ray powder diffraction (XRD) pattern of 
Sr(La)-112 before and after the grain alignment. After the alignment,
only the (002) reflection was seen in the XRD pattern. The inset of Fig.\ \ref{fig1}
shows the x-ray rocking curve of the (002) reflection. The full width at half maximum
of the reflection is less than 1 degree, which means a good $c$-axis alignment.
The aligned sample was approximately 9.5 mm in length and 3 mm in diameter.
The reversible magnetization was measured as a function of the temperature and 
the angle between the $c$ axis and the applied magnetic field by using a 
superconducting quantum interference device (SQUID) magnetometer 
(MPMS-XL, Quantum design).

\begin{figure}
\center
\includegraphics[width=7cm,height=5cm]{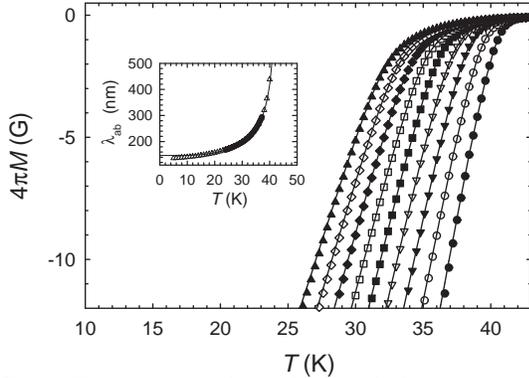}
\caption{Temperature dependence of the magnetization, $4\pi M(T)$, measured
         at applied magnetic fields of $1$ T $\leq H \leq$ 5 T 
         (filled circles,       1 T; open circles,        1.5 T;
         filled down triangles, 2 T; open down triangles, 2.5 T;
         filled squares,        3 T; open squares,        3.5 T;
         filled diamonds,       4 T; open diamonds,       4.5 T;
         filled up triangles,   5 T).
         The inset shows the penetration depth, $\lambda_{ab}(T)$, deduced
         from the Hao-Clem model (filled symbols) and the Shoenberg formula 
         (open symbols). In this figure, the solid line represents the BCS 
         temperature dependence of the penetration depth.}
\label{fig2}
\end{figure}

Figure\ \ref{fig2} shows the reversible magnetization, 4$\pi M(T)$, at fields of 
1 T $\leq H \leq$ 5 T parallel to the $c$ axis of the sample. In this figure,
the symbols and the lines denote the zero-field-cooled and the field-cooled
magnetizations, respectively. In comparison with 
other cuprates, our data show two interesting features. First, the curves shift to
lower temperature as the field increases and are almost parallel to each other.
This typical mean-field behavior is consistent with the prediction of
the Abrikosov model\cite{tinkham1} in which the magnetization scales linearly
with the magnetic field. Such a feature has never been observed before for any
high-$T_c$ cuprate because the mean-field behavior is usually screened by strong
thermal fluctuations.\cite{finnemore10} Secondly, the rate of decrease of $T_c(H)$ with respect to the
field is significantly larger than that of other cuprates, here the superconducting
transition temperature, $T_c(H)$, is estimated as the temperature at the point of
intersection of a linear extrapolation of $4\pi M(T)$ in the superconducting state
with the normal-state base line of $4\pi M=0$. For instance, at $H=5$ T, $T_c(H)$ 
is found to be about 33 K, which corresponds to $0.77T_c(0)$. Hence, the 
upper-critical field $H_{c2}(0)$ is expected to be about 20 T, assuming a linear 
$T_c(H)$. To determine $H_{c2}(0)$ precisely, we apply the Hao-Clem model for 
reversible magnetization\cite{hao1,hao2} to our data. The details of the analysis 
will be given elsewhere (M.-S. Kim {\em et al}., manuscript in preparation).
Two important parameters characterizing the compound are the Ginzburg-Landau parameter, 
$\kappa$, and the slope of the upper-critical field near $T_c$, $dH_{c2}/dT|_{T_c}$.
The Hao-Clem model analysis gives $\kappa=25.3\pm 1.1$ and 
$dH_{c2}/dT|_{T_c}=-0.47\pm 0.02$ T/K. Using these, we estimate 
$H_{c2}(0)$ to be $13.9\pm 0.5$ T through the relationship
$H_{c2}(0)\simeq0.7(dH_{c2}/dT)_{T_c}T_c$,\cite{werthamer1}
which is about ten times smaller than the value for other cuprate superconductors.
The in-plane coherence length $\xi_{ab}(0)=[\phi_0/2\pi H_{c2}(0)]^{1/2}$ is 
calculated to be $48.6\pm1.0$ \AA, where $\phi_0$ is the flux quantum.\cite{memo6}

In our study, we apply the Hao-Clem model to describe our magnetization 
data.\cite{mskim30,mskim21,mskim15}
Since the model is derived from the phenomenological Ginzburg-Landau (GL)
theory, our result is justified in the GL framework. However, we judge
our result not to be model dependent particularly from the following consideration:
The open symbols in the inset of Fig.\ \ref{fig2} represent the in-plane
magnetic penetration depth, $\lambda_{ab}(T)$, obtained from the dc-magnetic
susceptibility, $4\pi\chi(T)$, for the low-field region of $H<H_{c1}$. 
To deduce $\lambda_{ab}(T)$ from the $4\pi\chi(T)$ curve, we use
the Shoenberg formula,\cite{shoenberg1} which is not model dependent,
but is merely based on the London equations. For comparison, we also plot
$\lambda_{ab}(T)$ (filled symbols) from the Hao-Clem analysis in the same figure. 
We can see that the two curves, indeed, coincide. Thus, we can conclude that 
the application of the Hao-Clem model in this study does not reduce the generality 
of our results.

In the inset of Fig.\ \ref{fig2}, the solid line represents the temperature
dependence of the penetration depth assuming the BCS clean limit.\cite{tinkham1}
The estimated zero-temperature penetration depth, $\lambda_{ab}(0)$, 
is $147\pm6$ nm, which is close to $\lambda_{ab}(0)\simeq 130$ nm of Y-123. 
In the framework of the London model, the penetration depth is proportional to
$(m^\ast_{ab}/n_s)^{1/2}$, where $m^\ast$ and $n_s$ are the electronic effective
mass in the $ab$ plane and the charge-carrier density, respectively.
According to the empirical Uemura relation,\cite{uemura1} {\em i.e.}, 
$T_c\sim n_s/m^\ast_{ab}$, the $T_c$'s of the two compounds should be nearly 
the same. However, the $T_c$ of Sr(La)-112 is about half that of Y-123. 
This discordance might differentiate Sr(La)-112 from the hole-doped cuprates.

For the cuprate superconductors studied until now, the zero-temperature coherence
length along the $c$ axis, $\xi_c(0)$, was found to be much smaller than the
unit $c$-axis length. As the temperature increased toward $T_c$, 
a dimensional crossover from 2D to 3D occurred at a certain 
temperature $T^\ast$ where $\xi_c(T^\ast)=c/\sqrt{2}$.\cite{prober1} For moderately anisotropic 
materials like Y-123, a broader 3D-temperature region around $T_c$ was observed 
due to strong interlayer coupling.\cite{wclee1} However, the 3D region for strongly 
anisotropic compounds, such as Bi- and Tl-based superconductors, was found to be 
extremely narrow.\cite{li1}

Since the unit structure of an infinite-layer superconductor does not contain
a CRB, one can expect the coupling between CuO$_2$ planes to be strong.\cite{imai1}
To date, however, no rigorous studies to examine the dimensionality of the
compound have been done. With our high-quality samples, we designed experiments to 
obtain the anisotropy ratio $\gamma=\xi_{ab}/\xi_{c}$. A combination of
$\gamma$ and the $\xi_{ab}(0)$ obtained above gives the ratio $\xi_c/c$
which determines the dimensionality of the system.

Under an external magnetic field, the magnetization of all anisotropic materials 
has two components, $M_{\rm L}$ and $M_{\rm T}$. $M_{\rm L}$ is the component 
parallel to the field, and the other component, $M_{\rm T}$, is perpendicular 
to the field. Usually, $M_{\rm T}$ can be measured using a torque magnetometer, 
and the degree of anisotropy is reflected in the angular dependence of the torque,
$\tau(\theta)=HM_{\rm T}(\theta)$. Previously, Farrell {\em et al.}\cite{farrell2,farrell3} 
measured the magnetic torque curves for highly anisotropic 
Bi$_2$Sr$_2$CaCu$_2$O$_8$ (Bi-2212) and moderately anisotropic Y-123. By applying the 
London model to the data, they obtained the anisotropy ratios $\gamma\simeq55$ and 
$\gamma\simeq5$ for Bi-2212 and Y-123, respectively. On the other hand, we obtained 
the anisotropy ratio of Sr(La)-112 by measuring $M_{\rm L}(\theta)$ using a SQUID 
magnetometer with a sample rotator. The obtained data were analyzed by using 
the calculation of Hao {\em et al.}\cite{hao1,hao2} which considered the effective 
mass anisotropy of the material. For comparison, we also applied the London 
model\cite{tinkham1} to the data.

\begin{figure}
\center
\includegraphics[width=7cm,height=5cm]{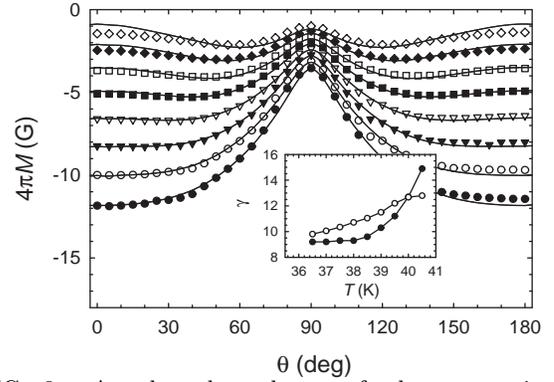}
\caption{Angular dependence of the magnetization, $4\pi M(\theta)$, measured
         at temperatures of 36.5 K $\leq T \leq$ 40 K  and a field of $H=1$ T 
         (filled circles,        36.5 K; open circles,        37 K;
         filled down triangles,  37.5 K; open down triangles, 38 K;
         filled squares,         38.5 K; open squares,        39 K;
         filled diamonds,        39.5 K; open diamonds,       40 K).
         The $\theta$ denotes the angle between the applied
         magnetic field and the crystallographic $c$ axis of the sample.
         The solid lines represent the prediction of the Hao-Clem model.
         The inset shows the anisotropy 
         ratio $\gamma (T)$ obtained from $4\pi M(\theta)$. The filled and the 
         open symbols are deduced from analyses based on the Hao-Clem and 
         the London models, respectively. The solid lines are just guides for the eyes.}
\label{fig3}
\end{figure}

Figure\ \ref{fig3} shows the angular dependence of the reversible magnetization,
$4\pi M(\theta)$, measured at $H=1$ T in the temperature range of 36.5 K $\leq T\leq$
40 K.\cite{memo3} In this figure, the solid lines represent the theoretical
prediction of the Hao-Clem model. We see a good fit, except for temperatures
above $T = 39$ K. We infer that the departure of the data from the
theoretical lines at higher temperatures is due to the thermal fluctuation effect,
which becomes more important as the temperature increases.\cite{mskim30,mskim14}
The inset of Fig.\ \ref{fig3} is a plot of the anisotropy ratio obtained from 
each curve in Fig.\ \ref{fig3}. The filled and the open symbols are deduced by 
application of the Hao-Clem and the London models, respectively. In the filled-symbol 
set, the curve shows a plateau behavior at low temperatures and then increases 
monotonically with temperature. The increase is postulated to originate from the 
thermal fluctuation effect, as mentioned above. In the open-symbol set, however, 
no such  plateau feature exists. In fact, the London model is known to be 
suitable for the low-field region of $H\ll H_{c2}$. Since our magnetization data 
were taken near the transition temperature, the external field of 1 T is regarded 
as considerably large. Thus, our data set is out of the London region. This is
consistent with the $4\pi M(T)$ data in Fig.\ \ref{fig2}, 
which lie in the Abrikosov region.

It is quite natural to take the value of the plateau in the filled-symbol set as 
the real anisotropy ratio. With this value, $\gamma=9.3\pm0.2$, we obtain an 
out-of-plane coherence length of $\xi_c(0)=5.2\pm0.3$ \AA~by using the relationship 
$\xi_{c}=\xi_{ab}/\gamma$ and the value of $\xi_{ab}(0)$ from the above analysis. 
The criterion for 3D superconductivity below $T_c$ is $\xi_c(T)>c/\sqrt{2}$.\cite{prober1}
The value of $\xi_{ab}(0)=5.4$ \AA~is considerably larger than the 
$c/\sqrt{2}\simeq2.4$ \AA. This means that the superconducting order 
parameter of one CuO$_{2}$ plane overlaps with those of neighboring CuO$_{2}$ 
planes even at zero temperature.

\begin{figure}[h]
\center
\includegraphics[width=7cm,height=5cm]{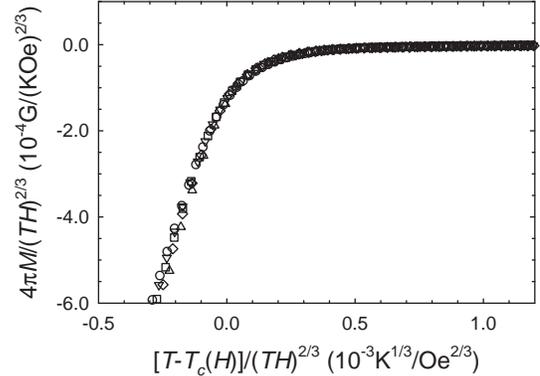}
\caption{Scaling of the data of Fig. 1 with the scaling variable 
         $(T-T_c(H))/(TH)^{2/3}$. In this analysis, a linear temperature dependence 
         of $H_{c2}$ near $T_c$ was used assumed, {\em i.e.}, $-dH_{c2}/dT = 0.47$ T/K.}
\label{fig4}
\end{figure}

Additional evidence for 3D superconductivity can be found from the scaling 
analysis\cite{mskim30,mskim21,mskim15,mskim14,li1,tesanovic2,jhsok1} 
of the fluctuation-induced magnetization for the high-field region. According to 
Ullah and Dorsey,\cite{ullah1} the magnetization in the critical region scales 
with the scaling variable of $A[T-T_c(H)]/(TH)^{n}$, where $A$ is a field and
transition temperature-independent coefficient, and $n$ is 2/3 for a 3D system
and 1/2 for a 2D system. As expected, the magnetization scales excellently with
the 3D form. Figure\ \ref{fig4} shows $4\pi M/(TH)^n$ versus the scaling
parameter $(T-T_c(H))/(TH)^n$ with $n=$ 2/3. All the data for the different
fields collapse onto a single curve. The slope $-dH_{c2}/dT\simeq 0.5$ T/K
near $T_c$ is obtained from this scaling analysis. This value is fairly
consistent with that deduced from the $c$-axis magnetization analysis.

Our results provide clear evidences for three-dimensional superconductivity,
even at zero temperature, in the infinite-layer compound 
Sr$_{0.9}$La$_{0.1}$CuO$_2$. Here, we deduced the anisotropy ratio $\gamma$ to 
be about 9 by measuring the angular dependence of the reversible magnetization. 
This value is somewhat larger than the value of $\gamma\simeq 5$ for Y-123. 
However, the absence of a CRB in the infinite-layer compound allows the order 
parameter of one layer to overlap with those of neighboring layers. In addition, 
we found that the empirical Uemura relation was not applicable to the case of
Sr$_{0.9}$La$_{0.1}$CuO$_2$. 

The authors thanks D. Pavuna, M. Sigrist, J. L. Tallon, P. M\"{u}ller, and N.-C. Yeh
for useful discussions. This work was supported by Creative Research Initiatives
of the Korean Ministry of Science and Technology.

\newpage

\end{multicols}

\begin{references}
\bibitem{siegrist1}
T. Siegrist, S.~M. Zahurak, D.~W. Murphy, and R.~S. Roth, Nature {\bf 334},
  231  (1988).

\bibitem{smith1}
M.~G. Smith {\it et~al.}, Nature {\bf 351},  549  (1991).

\bibitem{jorgensen1}
J.~D. Jorgensen {\it et~al.}, Phys. Rev. B {\bf 47},  14654  (1993).

\bibitem{er1}
G. Er, S. Kikkawa, M. Takahashi, and F. Kanamaru, Physica C {\bf 276},  315
  (1987).

\bibitem{jones1}
E.~C. Jones, D.~P. Norton, D.~K. Christen, and D.~H. Lowndes, Phys. Rev. Lett.
  {\bf 73},  166  (1994).

\bibitem{liu1}
R.-S. Liu {\it et~al.}, Solid State Comm. {\bf 118},  367  (2001).

\bibitem{cujung1}
C.~U. Jung {\it et~al.}, Physica C (to be published).

\bibitem{kitazawa1}
K. Kitazawa, Physica C {\bf 341-348},  19  (2000).

\bibitem{Farrell1}
D.~E. Farrell {\it et~al.}, Phys. Rev. B {\bf 36},  4025  (1987).

\bibitem{mskim30}
M.-S. Kim, C.~U. Jung, S.-I. Lee, and A. Iyo, Phys. Rev. B {\bf 63},  134513
  (2001).

\bibitem{mskim21}
J.-H. Choi {\it et~al.}, Phys. Rev. B {\bf 58},  538  (1998).

\bibitem{mskim15}
M.-S. Kim {\it et~al.}, Phys. Rev. B {\bf 57},  8667  (1998).

\bibitem{mskim14}
M.-S. Kim {\it et~al.}, Phys. Rev. B {\bf 57},  6121  (1998).

\bibitem{tinkham1}
M. Tinkham, {\em Introduction to Superconductivity}, 2nd ed. (McGraw-Hill, New
  York, 1996).

\bibitem{finnemore10}
D.~K. Finnemore,  in {\em Phenomenology and Applications of High-Temperature
  Superconductors}, edited by K.~S. Bedell {\it et~al.} (Addison-Wesley,
  Massachusetts, 1992), p.\ 164.

\bibitem{hao1}
Z. Hao and J.~R. Clem, Phys. Rev. Lett. {\bf 67},  2371  (1991).

\bibitem{hao2}
Z. Hao {\it et~al.}, Phys. Rev. B {\bf 43},  2844  (1991).

\bibitem{werthamer1}
N.~R. Werthamer, E. Helfand, and P.~C. Hohenberg, Phys. Rev. {\bf 147},  295
  (1966).

\bibitem{memo6}
One of interesting points is that the superconducting parameters such as $T_c$,
  $\kappa$, $\xi_{ab}(0)$, $\lambda_{ab}(0)$, and the critical fields obtained
  in this study are the same as those of the newly discovered MgB$_2$
  superconductor within 10\% error range. (D. K. Finnemore {\em et al.}, Phys.
  Rev. Lett. {\bf86}, 2420 (2001)).

\bibitem{shoenberg1}
D. Shoenberg, {\em Superconductivity} (Cambridge University, Cambridge, 1954).

\bibitem{uemura1}
Y.~J. Uemura {\it et~al.}, Phys. Rev. Lett. {\bf 66},  2665  (1991).

\bibitem{prober1}
D.~E. Prober, M.~R. Beasley, and R.~E. Schwall, Phys. Rev. B {\bf 15},  5245
  (1977).

\bibitem{wclee1}
W.~C. Lee, R.~A. Klemm, and D.~C. Johnston, Phys. Rev. Lett. {\bf 63},  1012
  (1989).

\bibitem{li1}
Q. Li, M. Suenaga, T. Hikata, and K. Sato, Phys. Rev. B {\bf 46},  5857
  (1992).

\bibitem{imai1}
T. Imai, C.~P. Slichter, J.~L. Cobb, and J.~T. Markert, J. Phys. Chem. Solids
  {\bf 56},  1921  (1995).

\bibitem{farrell2}
D.~E. Farrell {\it et~al.}, Phys. Rev. Lett. {\bf 63},  782  (1989).

\bibitem{farrell3}
D.~E. Farrell {\it et~al.}, Phys. Rev. Lett. {\bf 61},  2805  (1988).

\bibitem{memo3}
Below $T\simeq 36$ K, the angular dependence of magnetization is severely
  asymmetric with respect to the angle $\theta=\pi/2$, which means that the
  sample does not enter an reversible state for entire region of $\theta$, {\em
  i.e.}, $0 \leq \theta \leq \pi$.

\bibitem{tesanovic2}
Z. Te\v{s}anovi\'{c} {\it et~al.}, Phys. Rev. Lett. {\bf 69},  3563  (1992).

\bibitem{jhsok1}
J. Sok {\it et~al.}, Phys. Rev. B {\bf 51},  6035  (1995).

\bibitem{ullah1}
S. Ullah and A.~T. Dorsey, Phys. Rev. Lett. {\bf 65},  2066  (1990).
\end{references}
\end{document}